\documentclass[%
 aip,
 jmp,%
 amsmath,amssymb,
 reprint,%
author-numerical,%
nofootinbib,superscriptaddress]{revtex4-1}

\usepackage{graphicx}
\usepackage{dcolumn}
\usepackage{bm}
\usepackage{subfigure}
\usepackage{color} 
\renewcommand{\thefootnote}{\fnsymbol{footnote}}
\begin{document}

\preprint{AIP/123-QED}

\title[]{Amplification of a surface electromagnetic wave by running over plasma surface ultrarelativistic electron bunch as a new scheme  for generation of Teraherz radiation} 
\thanks{Here we consider the spatially unbounded plasma-(long)bunch systems which can also be referred to as the spatially unbounded  plasma-bunch systems.}
\author{S. P. Sadykova}%
\altaffiliation{Electronic mail: \textcolor{blue}{Corresponding author - saltanat@physik.hu-berlin.de.}}
\affiliation{Max-Born Institut f\"ur Nichtlineare Optik und Kurzzeitspektroskopie im Forschungsverbund Berlin e.V.,  Max-Born-Stra\ss e  2 A,  Berlin, 12489, Germany
}
\author{A. A. Rukhadze}
\email{rukh@fpl.gpi.ru}
\affiliation{Prokhorov General Physics Institute,
Russian Academy of Sciences, Vavilov Str., 38., Moscow, 119991, Russia}

\author{T. G. Samkharadze}
\noaffiliation
\affiliation{Prokhorov General Physics Institute,
Russian Academy of Sciences, Vavilov Str., 38., Moscow, 119991, Russia}

\author{K. V. Khishchenko}
\affiliation{Joint Institute for High Temperatures RAS, Izhorskaya 13 bldg 2, Moscow 125412, Russia} 

\date{\today}

\begin{abstract}
The amplification of a surface electromagnetic wave by means of ultrarelativistic monoenergetic electron bunch running over the flat plasma surface in absence of a magnetic field is studied theoretically. It is shown that when the ratio of electron bunch number density to plasma electron number density multiplied by a powered to 5 relativity factor is much higher than 1, i.e $\gamma^5 n_b/n_p>> 1$,  the saturation field of the surface electromagnetic wave induced by trapping of bunch electrons gains the magnitude: $E_x=B_y\approx 0.16 \frac{\omega_p m c}{e} (\frac{2n_b}{\gamma^2 n_p})^{1/7}$ and does not approache the surface electromagnetic wave  front breakdown threshold in plasma. The surface electromagnetic wave saturation energy density in plasma can exceed the electron bunch energy density. Here, we discuss the possibility of generation of superpower Teraherz radiation on a basis of such scheme.  %

\end{abstract}

\maketitle

\section{\label{sec:Int}Introduction}

The surface electromagnetic waves (SEW) on plasma surface and plasma-like media (gaseous plasma, dielectric and conducting media, etc.) attract special attention of researchers due to their unique properties. First of all, due to its high phase and group velocities close to light speed in vacuum at high media conductivity what makes them the most valuable in radiophysics \cite{5}. The SEW are widely applied in physical electronics due to its high phase velocity leading to its uncomplicated generation by relativistic electron bunches and   output from plasma.\\
\indent Below we discuss the problem of SEW amplification with the help of electron bunch running over flat plasma surface. We consider  the case of ultrarelativistic monoenergetic electron bunch which remains relativistic in the frame of reference of SEW generated by this  bunch compared to the works \cite{1,7,8}, where the bunches were nonrelativistic. Such a problem of generation of three-dimensional electromagnetic wave (wakefields) in plasma with the help of ultrarelativistic electron and ion bunches through Cherenkov resonance  radiation was solved in \cite{2}, where it was shown that bunch ultrarelativity influences significantly the nonlinear stage of  plasma-bunch interaction, in particular, the saturation amplitude of the  generated wave.   \\ 
\indent In the present work we apply the method developed in \cite{2} for the case of amplification of a surface electromagnetic wave by means of ultrarelativistic monoenergetic electron bunch running over the flat plasma surface. The interest to the SEW amplification was aroused by its uncomplicated output from plasma compared to that of the three-dimensional wave generated by the bunch as well and high  magnitudes of SEW energy density. The latter is related to the field structure. Thus, as it'll be shown below, the SEW saturation energy density can exceed the bunch energy density. \\
\indent It is noteworthy that the real SEW amplification device should be cylindrical what we do comprehend very well. However, the problem taking into account the cylindrical geometry is much more complex compared to that of plane geometry from the mathematical point of view and is not appropriate for illustrative purposes. This is why we restrict ourselves to the plane geometry problem. Soon, we are planning to finish an article considering the real cylindrical SEW bunch-plasma amplifier and will present it for publication.

	\section{Description of the Model. Dispersion Relation}
Let us start our description with the schematic illustration of interaction of the ultrarelativistic monoenergetic electron bunch with cold isotropic plasma (no thermal motion) being in a rest, which generates the plane wave $E=E_0 \exp(-i\omega t + i \vec k \cdot \vec r)$, and put the external field as absent. \protect \footnote{In ultrarelativistic bunches bunch divergence can be neglected because when the electron bunch is ejected into the dense plasma, within the time $t\sim 1/\omega_p$ the neutralization of the bunch charge occurs prohibitting  the bunch divergence}. 

	\begin{figure*}
{\includegraphics[height=6.8 cm,width=7.6cm]{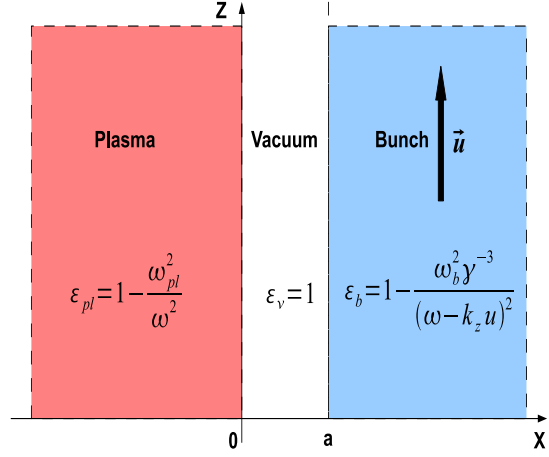}}
\caption{ Schematic illustration of interaction of the ultrarelativistic monoenergetic electron bunch with plasma. Here, the corresponding dielectric permitivities are presented as well.}
\label{Fig:1}
\end{figure*}
Over the colissionless plasma, filling in the half-plane $x<0$, with the dielectric permitivity
 \begin{equation}
\label{1}
\varepsilon_p=1-\frac{\omega_p^2}{\omega^2},
\end{equation} 
the ultrarelativistic monoenergetic electron bunch, filling in the space $x\geq a$, with the dielectric permitivity
\begin{equation}
\label{2}
\varepsilon_b=1-\frac{\omega_b^2\gamma^{-3}}{(\omega-k_z u)^2},
\end{equation} 
propogates on a distance $a$. Here $\omega_p= \sqrt{4\pi e^2 n_p/m}$, $\omega_b= \sqrt{4\pi e^2 n_b/m}$ are Langmuir plasma electron and bunch frequencies  respectively (in GSU units) with $n_p$, $n_b$ being the plasma and bunch number densities in the laboratory frame of reference (plasma in a rest) ($n_b\ll n_p$), $k_z$ is the longitudinal (directed along the velocity of the bunch $\vec u$) component  of the SEW wave vector  $\vec k=(k_x,0,k_z)$, $e$= the electron charge, $m$= its mass. 
The bunch is considered to be an ultrarelativistic when
\renewcommand{\thefootnote}{\arabic{footnote}}\setcounter{footnote}{0}
\begin{equation}
\label{3}
 \gamma = \frac{1}{\sqrt {1-u^2/c^2} } >> 1.
\end{equation} 
\indent The surface wave is a wave of $E$-type with the nonzero field components $E_x,E_z, B_y$, which satisfy the following system of equations \cite{8}:
\begin{equation}\label{4}
\begin{gathered}
\frac{\partial^2 E_z}{\partial x^2}-{k_z}^2 E_z+\frac{\omega^2}{c^2}\varepsilon(x)E_z=0 \hfill \\
E_x=-\frac{\imath k_z}{\kappa^2}\frac{\partial E_z}{\partial x}, \:\: B_y=-\frac{i\omega}{c \kappa^2}\varepsilon(x)\frac{\partial E_z}{\partial x},
\end{gathered}
\end{equation}
where $\kappa^2={k_z}^2-\varepsilon\omega^2/{c^2}$. The system (\ref{4}) is valid for all domains shown in Fig. \ref{Fig:1} with the corresponding substitutions $\varepsilon=\varepsilon_p$, $\varepsilon=\varepsilon_\nu=1$, $\varepsilon=\varepsilon_b$. The electric fields are the following functions of the time and the coordinates 
\begin{equation}
\label{5}
E_z=E_{0z}(x) \exp(-i\omega t + i k_z z).
\end{equation}
Dependence on $x$ is defined by the system (\ref{4}) and can be represented as follows
\begin{equation}\label{6}
E_{0z}(x) = \left\{   \begin{array}[c]{ll}
 C_1 e^{\kappa_p x} &\mbox{at} \: x\leq 0,\nonumber \\
 C_2 e^{\kappa_\nu x}+C_3 e^{-\kappa_\nu x}& \mbox{at}\: 0\leq x\leq a \nonumber, \\
 C_5 e^{-\kappa_b x} & \mbox{at}\: x\geq a \nonumber, 
\end{array} \right. \end{equation}
where $\kappa_p=\sqrt{k_z^2-\omega^2\varepsilon_p/c^2}$, $\kappa_\nu=\sqrt{k_z^2-\omega^2/c^2}$ and $\kappa_b=\sqrt{k_z^2-\omega^2\varepsilon_b/c^2 }$.\\
\indent The boundary conditions can be obtained from the field equations by integrating over a thin layer near the interface between two corresponding media and have the following view:
\begin{equation}
\label{7}
\begin{gathered}
{E_z}_p\vert_{x=0}={E_z}_\nu\vert_{x=0}, \; {E_z}_\nu\vert_{x=a}={E_z}_b\vert_{x=a}, \hfill \\
 {B_y}_p\vert_{x=0}={B_y}_\nu\vert_{x=0},\;{B_y}_\nu\vert_{x=a}={B_y}_b\vert_{x=a}.
\end{gathered}
\end{equation}
In addition to these boundary conditions the following condition must be satisfied:
\[
\begin{gathered}
{E_z}_p\vert_{x\to \pm\infty}={E_z}_b\vert_{\to \pm\infty}=0,  \hfill \\
 {B_y}_p\vert_{x\to \pm\infty}={B_y}_b\vert_{\to \pm\infty}=0.
\end{gathered}
\]
\indent Having solved the system of equations (\ref{4})- (\ref{7}) we can finally obtain the following dispersion relation:   
\begin{equation}
\label{8}
\varepsilon_p\kappa_\nu+\kappa_p=-\frac{\kappa_p}{2} \exp(-2a\kappa_\nu)(\varepsilon_b-1)(1+\frac{{k_z}^2}{{\kappa_\nu}^2}).
\end{equation}
\indent When the bunch is absent, i.e. $n_b=0$ and $\varepsilon_b=1$, one can get the dispersion relation of surface plasma wave from the following equation:
\begin{equation}
\label{9}
\varepsilon_p\sqrt{k_z^2-\omega^2/c^2}+\sqrt{k_z^2-\omega^2\varepsilon_p/c^2 }=0,
\end{equation}
  which was studied with the solution $\omega=\omega_0$ in detail in \cite{5,8}. The bunch leads to the amplification of this wave and   solution of Eq. (\ref{7}) should be found in the following form:
\begin{equation}\label{10}
\omega=k_z u(1+\delta)=\omega_0(1+\delta), \; \; \delta\ll 1.
\end{equation}   
Since we took into account that $n_b\ll n_p$, the highest bunch effect on the surface wave occurs when the following Cherenkov resonance  condition is satisfied
\begin{equation}\label{11}
\omega_0=k_z u.
\end{equation}   
\section{Analysis of the dispersion relation. Bunch-assisted generation of the surface electromagnetic wave under the Cherenkov resonance  condition.}
\begin{figure*}
{\includegraphics[height=8 cm,width=12cm]{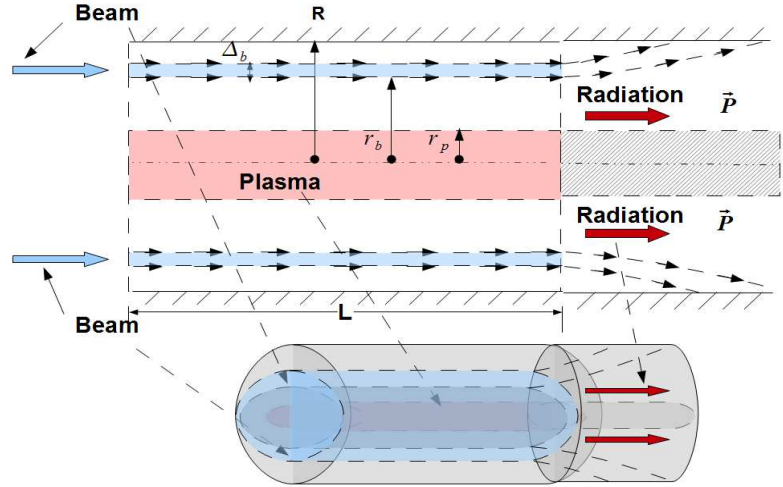}}
\caption{Schematic drawing of surface electromagnetic wave amplifier. Here, $r_b>>\Delta_b$ and $c/\omega_p<< r_p$, i. e. the plasma surface can be considered as a flat one.}
\label{Fig:2}
\end{figure*}

Let us first determine the SEW frequency in a bunch absence, i.e. find solution  of Eq. (\ref{4}). We are interested in the frequency range of high-speed waves with $\omega_0\approx k_z c$ which can be generated by an ultrarelativistic bunch under Cherenkov resonance  condition, i.e. $\omega_0\approx k_z u$. From Eq. (\ref{9}) follows that such waves can exist only in dense plasmas when $\omega\ll \omega_p$ and hence $\varepsilon_p\approx - \omega_p^2/\omega^2$. From Eq. (\ref{9}) we can easily find 
\begin{equation}\label{12}
\omega_0^2=\frac{\omega_p^2}{\gamma^2+1}\approx \frac{\omega_p^2}{\gamma^2}, 
\end{equation}   
where the inequality (\ref{3}) was taken into account.\\
\indent Let us now take into account the bunch effect, i.e find solution of Eq. (\ref{8}) when the Cherenkov resonance  condition (\ref{11}) is satisfied. Here, we would like to restrict ourselves to consideration of the ultrarelativistic bunch, when $|\delta|\gamma^2 >> 1$, running over and closely to the flat plasma surface, $2^{3/2}a\omega_0 \sqrt{\delta}/c<< 1$, for this - $\exp(-2a\kappa_\nu)\simeq 1$. Then one can get the following solution for $\delta$ 
\begin{equation}\label{13}
\delta^{7/2}=\frac{\omega_b^2}{4\sqrt{2}\omega_p^2\gamma^2}.
\end{equation}   
The solution of Eq. (\ref{13}), we are interested in, has the following form
\begin{equation}\label{14}
\delta=\delta'+i\delta''=\left(\cos\left(\frac{6\pi}{7}\right)+i \sin\left(\frac{6\pi}{7}\right) \right)\left(\frac{n_b}{4\sqrt{2}n_p\gamma^2}\right)^{2/7}.
\end{equation}   
It is obvious that the saturation of instability can occur when the kinetic energy of electrons, in the SEW frame of reference, will become less than the amplitude of the potential of the plasma wave measured in the same frame. In this case the bunch electrons get trapped by the SEW, i.e. there will be no relative motion between the bunch electrons and the SEW, thus, no energy exchange between the bunch and SEW occurs, the bunch and the SEW become stationary. For determination of the saturation amplitude of the potential of the plasma SEW, generated by the bunch and amplified with the time with the increment $\Im m\delta=\delta''$, we will apply the same method used in \cite{2}. Let us choose the SEW  frame of reference in which the wave is purely potential and its stationary saturation amplitude $\Phi'$ can be determined by condition of the bunch electrons trapping  in the wave field \cite{2}. 
 \begin{equation}
   \label{15}
	\frac{e \Phi'}{mc^2} =\frac{1}{\sqrt{1-\frac{{u_1}^2 }{c^2}}}-1   ,
	\end{equation}
	where $\Phi'$ is the SEW potential and $u_1$ is the bunch electrons velocity, both measured in the SEW frame of reference.  
	In accordance with Lorentz transformations the speed of bunch electrons in the chosen frame will be 
 \begin{equation}
   \label{16}
   \begin{gathered} 
 u_1=-\frac{u \delta' \gamma^2}{1- \frac{2u^2}{c^2} \delta' \gamma^2}=- \frac{u  \delta' \gamma^2}{1- 2 \delta' (\gamma^2-1)}=\hfill \\ \left\{\begin{array}[c]{ll}  u_1 \approx -u \delta' \gamma^2<< u, &\mbox{at} 2|\delta'|\gamma^2 << 1\nonumber \\
  u_1 \approx u/2, &\mbox{at} (2|\delta'|\gamma^2 >> 1),
  \end{array} \right.
 \end{gathered}
 \end{equation}  
  here the real part of $\delta$ ($\delta'$) is considered.\\
 \indent In the laboratory frame of reference the potentials $\Phi_0$ and $A_z$ are not zero and
 \begin{equation}
   \label{17}
\Phi_0= \Phi' \gamma, \; \;	A_z=\frac{u}{c}\Phi_0\simeq \gamma\Phi' .
	\end{equation}
Knowing $\Phi_0$ and $A_z$ we can determine the fields in the laboratory frame of reference:
\begin{equation}\label{18}
 \begin{array}[c]{l} 
E_z=-\frac{1}{c}\frac{\partial A_z}{\partial t} - \frac{\partial\Phi_0}{\partial z}=-i k_z \Phi_0+ i\frac{\omega}{c} A_z=-i k_z \gamma \Phi' (1-\frac{\omega u}{k_z c^2}),  \\
E_x=-\frac{\partial \Phi_0}{\partial x}, \; \; B_y=-\frac{\partial A_z}{\partial x}\simeq E_x  .
 \end{array}
\end{equation}
It is obvious that the amplified by the bunch SEW wave can be easily radiated out of the vacuum domain, in which the bunch is running (see Fig. \ref{Fig:2}). At the same time, the transverse fields  $E_x$  and $B_y$ are much higher than the longitudinal fields. This is  why we restrict ourselves to calculation of the transverse fields in the vacuum domain, also because  these components form the much higher longitudinal radiating component of Poynting vector (energy density flux) compared to the transverse one.
 \begin{equation}
   \label{19}
P= \frac{c}{4\pi} E_x B_y=\frac{c}{4\pi} {E_x}^2
	\end{equation}
It is quite easy to calculate the fields $E_x$  and $B_y$ in the plasma domain from Eq. (\ref{18}). It is noteworthy that in our previous work \cite{2} the electric field generated by the  ultrarelativistic bunch in the range $|\delta|\gamma^2 >> 1$ was decreasing with an increase of $\gamma$ and we had to determine the $\gamma$ at which the field is the highest. In this case, the behaviour is similar: they increase with $\gamma$ at $|\delta|\gamma^2 << 1$ and decrease in the range $|\delta|\gamma^2 >> 1$. In the considered case the ultrarelativistic range $|\delta|\gamma^2 >> 1$ is of high interest. In this range we can obtain the following fields: 
\begin{equation}\label{20}
\begin{gathered}
\vert E_x \vert=\vert B_y \vert=\vert-\frac{\partial \Phi_0}{\partial x}\vert=\vert -i \kappa_\nu \Phi_0\vert \hfil \\  =\frac{\omega_p}{u}\sqrt{2\vert \delta \vert}\Phi'=0.16 \frac{\omega_p m c}{e} (\frac{2n_b}{\gamma^2 n_p})^{1/7}. \hfil
\end{gathered}
\end{equation}
These fields are lower than the maximum stationary (saturation) plasma wave field obtained in \cite{3} being equal to $E_{max}=\sqrt{2}\omega_p m c\sqrt{\gamma}/e$. It is assumed that at this magnitude the surface electromagnetic wave front breakdown occurs. In the considered case the transverse SEW is generated and its behaviour is not trivial. The only thing we can say is that the fields (\ref{20}) are determined from condition of the bunch electrons trapping 
in the wave field and that they are lower than those obtained from SEW front breakdown threshold. \\
 \indent Finally, let us determine the SEW saturation energy density with the corresponding amplitudes (\ref{20})
  \begin{equation}\label{21}
W=\frac{1}{8\pi}({E_x}^2+{B_y}^2)=\frac{1}{4\pi}{E_x}^2=0.024 m c^2 \left(\frac{2n_b n_p^{5/2}}{\gamma^2}\right)^{2/7}.
\end{equation}
These magnitudes can exceed the bunch electrons energy density $E_b \simeq mc^2 \gamma n_b$ and this fact should not surprise a reader (see the Results and Discussions).
 
\section{Results and Discussions}
\indent Let us begin with the concluding words of the last section: The saturation energy density of surface electromagnetic wave  (\ref{21}),  generated by means of ultrarelativistic electron bunch running over the flat plasma surface, can exceed the bunch electrons  energy  density and this fact should not surprise a reader. This is due to the fields structure, i. e. $\vert E_x\vert =\vert B_y\vert>> \vert E_z \vert$. The point is that the bunch gets trapped by SEW weak longitudinal field component $E_z$, whereas the bunch energy is transfered to the whole SEW, i.e. the transverse components $E_x$ and $B_y$ get amplified as well forming the longitudinal energy density flux.  As a result, the SEW  saturation energy density becomes considerably higher than the corresponding bunch energy density. \\
\indent  The schematic view of the real SEW amplifier is presented in Fig. 2. Obviously, the accelerator represents a plasma cylinder (plasma is generated in the glass cylinder with the given gas pressure) of fixed length $L$ which is blowed around by the  ultrarelativistic bunch. In the plasma cylinder of radius $r_p <r_b$ (bunch radius) plasma of given number density $n_p$ is generated. The  ultarrelativstic bunch of radius $r_b>>\Delta_b$ (bunch width) and of given number density $n_b$, the current $J_b=2\pi r_b \Delta_b e n_b u$,  penetrates the metallic vacuumed cylindrical chamber of radius $R$ and length $L$ from the left end and comes out at the right end   being detected by a bunch detector.  To the right end of the plasma chamber a metallic coaxial chamber is docked where into the SEW, transformed into the coaxial transverse electric and magnetic mode, is let out. At the junction a partial SEW (longitudinal  component) reflection occurs and a quasistatic wave with increasing along Z-axis amplitude gets formed.  Only running toward Z-axis wave (forward wave) interacts with the bunch whereas the backward wave - does not. \\
\indent In conclusion let us make some estimations pursuing the goal of employment of the presented above model for constructing of  superpowerful Teraherz radiator ($f_0\simeq 0.5 \cdot 10^{12}$ Hz, $\omega_0=3\cdot 10^{12}$ s$^{-1}$). Since $\omega_0=\omega_p/\gamma$ then the plasma and bunch parameters can be chosen respectively. Today, the high-current accelerators are the linear accelerators with electron energy of $50$ MeV ($\gamma=100$)  and current density of $500$ A/cm$^{-2}$ ($n_b=10^{11}$ cm$^{-3}$, net current $I=3$ A). The plasma frequency should be of order $\omega_p=\gamma \omega_0=3\cdot 10^{14}$ s$^{-1}$ and $n_p\simeq 3 \cdot 10^{19}$ cm$^{-3}$, atmospheric pressure. Correspondingly, the time increment will  be $\omega_0\delta''\simeq 2\cdot 10^8$ s$^{-1}<<\omega_0$ at $a=0.1$ and the amplification coefficient $\delta'' k_z\simeq 1/L\simeq 7.4\cdot 10^{-3}$ cm$^{-1}$ leading to the system length of 1.34 m where the plasma radius is $r_p=0.1 $ cm and beam radius is $r_b=0.2$ cm; $c/\omega_p=10^{-4}<<r_p$, i.e. the plasma surface can be considered as a flat one.  Let us notice that the condition $\gamma^2\delta'\simeq 3 > 1$ is satisfied. Finally, let us estimate the SEW radiation flux. From (\ref{20}) follows that $\vert E_x \vert\simeq \vert B_y \vert\simeq 10^{7}$ V/cm=$ 10^{9}$ V/m, hence, the  Poynting vector $\vert P \vert=c/4\pi (E_x^2)\simeq 6 \cdot 10^{11}$ W/cm$^{-2}$.    \\
\indent Finally, let us notice that presented above results are valid only for collisionless plasma, i. e. $\omega_0>>\nu_e$ with $\nu_e$ being the collisions frequency, satisfying Eq. (\ref{1}). For plasma of electron number density $n_p$ and temperature $T_e$ we have $\nu_e\sim \frac{10 n_p}{{T_e}^{3/2}}$ , where $T_e$ is measured in Kelvin \cite{8}. At $\omega_0=3 \cdot 10^{12}$ s$^{-1}$ and $n_p=3 \cdot 10^{19}$ cm$^{-3}$ the condition $\omega_0>>\nu_e$ is satisfied when $T_e> 2\cdot 10^5$ K. Such temperature can be gained during the ionization of a gas of atmospheric pressure only with the help of powerful lasers.
\appendix

\begin{acknowledgments}
 S.P. Sadykova would like to express her gratitude to her father P. S. Sadykov for his financial support of the work and for being all the way the great moral support for her.
\end{acknowledgments}

\nocite{*}

\end{document}